\documentclass[3p,times,twocolumn,preprint]{elsarticle}
\journal{arXiv}
\usepackage{graphicx}
\usepackage{subfig}
\usepackage[colorlinks=true]{hyperref}
\usepackage{amsmath}
\usepackage{amssymb}
\begin{document}
\begin{frontmatter}
\title{Ferroelectric Control of Metal-Insulator Transition}
\author[iphy]{Xu He}
\author[iphy,cic]{Kui-juan Jin\corref{cor1}}
\ead{kjjin@iphy.ac.cn}
\author[iphy]{Chen Ge}
\author[cic,pku]{Zhong-shui Ma}
\author[iphy,cic]{Guo-zhen Yang}

\address[iphy]{Beijing National Laboratory for Condensed Matter Physics,\\ Institute of Physics, Chinese Academy of Sciences, Beijing 100190, China}
\address[cic]{Collaborative Innovation Center of Quantum Matter, Beijing 100190, China}
\address[pku]{School of Physics, Peking University, Beijing 100871, China}
\cortext[cor1]{Corresponding author}

\begin{abstract}
We propose a method of controlling the metal-insulator transition of one perovskite material at its interface with a another ferroelectric material based on first principle calculations. The operating principle is that the rotation of oxygen octahedra tuned by the ferroelectric polarization can modulate the superexchange interaction in this perovskite. We designed a tri-color superlattice of (BiFeO$_3$)$_N$/LaNiO$_3$/LaTiO$_3$, in which the BiFeO$_3$ layers are ferroelectric, the LaNiO$_3$ layer is the layer of which the electronic structure is to be tuned, and LaTiO$_3$ layer is inserted to enhance the inversion asymmetry. By reversing the ferroelectric polarization in this structure, there is a metal-insulator transition of the LaNiO$_3$ layer because of the changes of crystal field splitting of the Ni $e_g$ orbitals and the bandwidth of the Ni in-plane $e_g$ orbital. It is highly expected that a metal-transition can be realized by designing the structures at the interfaces for more materials.
\end{abstract}
\end{frontmatter}
\section{Introduction}
In transitional metal oxides,the strong correlation between lattice, charge, orbital, and spin leads to many novel properties. At their interfaces, even richer physics bring about emerging properties. Technical advances in the atomic-scale synthesis of oxide heterostructures make it possible for these interfaces to be artificially designed~\cite{Mannhart26032010,hwang2012emergent,chakhalian2014colloquium}. In perovskite oxide heterostructures where the oxygen octahedra share their vertices, the interplay among different distortion of octahedral units can dictate many novel functional properties~\cite{MRS:8511543}. One kind of interfaces between ferroelectric and other materials, of which the electronic properties are to be controlled, is very interesting because of the bi-stable property, with which two states can be reached by reversing the ferroelectric polarization with an electric field, with both changes from structure and from electric polarization involved. The controlling of the electronic structure can be through either the change of electric boundary condition~\cite{zubko2011interface}, or structural distortion~\cite{PhysRevLett.105.087204}. In this paper, we propose a strategy to control the electronic bandwidths and to achieve a metal-insulator transition in a material by stacking it onto a ferroelectric layer resulting in a modulation of structural distortion.

In this work, we use LaNiO$_3$ as the electronic active material, of which the band gap is to be controlled. Nickelates, with the chemical formula RNiO$_3$, have a prominent feature that there is a metal-insulator transition connected to the sizes of the R site ions by the rotation angles of octahedra~\cite{medarde1997structural}. The rotation of the octahedra can also be controlled by a strain, causing many works on the straining control of the electronic structure of nickelates such as in Refs. ~\cite{PhysRevB.82.014110,PhysRevB.88.195108,PhysRevB.87.060101}. In addition, nickelate heterostructures have drawn great attentions since Chaloupka and Khaliullin proposed the possible superconductivity by modulating the orbitals~\cite{PhysRevLett.100.016404}.  With the proposed structure, the orbital occupations can also be tuned. We chose BiFeO$_3$ as the ferroelectric material because of its large polarization. It has a relative small band gap, which enables it be used as semiconductor materials such as in switchable diodes\cite{choi2009switchable,Jiang2011Resistive,wang2011switchable,ge2011numerical}, therefore it is easier for charges to transfer from or to BiFeO$_3$. One can make use of this effect to manipulate the charge transfer.

By stacking a ferroelectric and non-ferroelectric layers together, it is natural to think that the electronic properties can be tuned by reversing the ferroelectric polarization. While it is true for the asymmetric thin films, in a periodic superlattice, the structures of the two polarization states are the same (or very similar) by a 180 degree rotation if the two interfaces of a non-ferroelectric to the ferroelectric are the same (or very similar). For example, in a (BiFeO$_3$)$_N$/LaNiO$_3$ superlattice, the two sides of LaNiO$_3$ are the LaO and BiO planes. Since the La and Bi ions are close in radius and the same in valence states, large difference in electronic properties caused by reversing the polarization are not expected.

 A method to enlarge the difference is to make the structure more asymmetric by inserting another layer. By inserting a LaTiO$_3$ layer,tri-color superlattices of (BiFeO$_3$)$_N$/LaNiO$_3$/LaTiO$_3$ is formed as shown in Fig.~\ref{fig:struct}. According to Chen, \emph{et al.}~\cite{chen2013modifying}, in LaTiO$_3$, the Ti$^{3+}$ ion has a $3d$ electron with energy higher than that of the unoccupied Ni $e_g$ band in LaNiO$_3$. Therefore electrons can transfer from Ti to Ni, forming a Ti $d^0$ and Ni $d^8$ configuration~\cite{karolak2015nickel}, so there is an electric polarization pointing from LaNiO$_3$ to LaTiO$_3$ at their interface. Thus the structures with opposite ferroelectric polarizations also differ in polarization continuity, which offers another way of controlling the electronic property by reversing the ferroelectric polarization. In the LaNiO$_3$/LaTiO$_3$ heterostructures, the crystal field splitting of the Ni $e_g$ orbitals is largely affected by the distortion of the oxygen octahedra caused by the polarity of the structure~\cite{chen2013modifying}. Thus the control of the ferroelectric polarization in the tri-color superlattice can also tune the electronic structure by influence the octahedra distortion.
\section{Methods}
\begin{figure}[htb]
  \centering
  \includegraphics[width=0.48\textwidth]{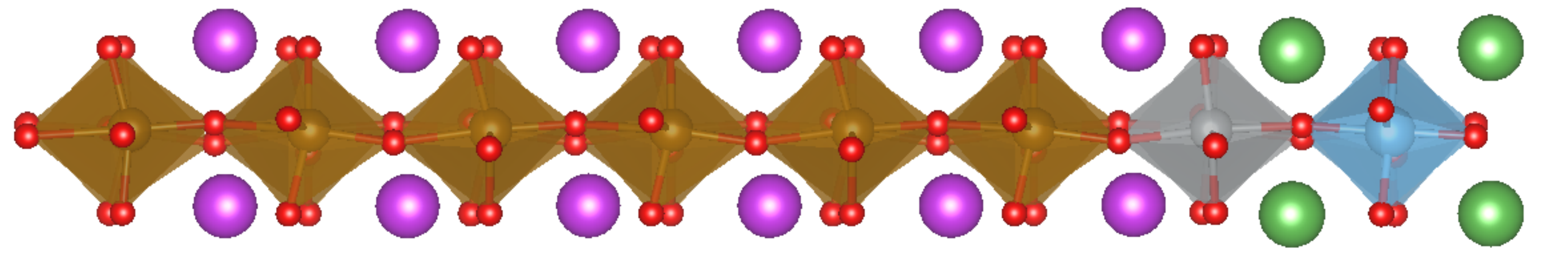}
  \caption{The structure of the BiFeO$_3$/LaNiO$_3$/LaTiO$_3$ superlattice in the $ac$ plane view. The purple, brown, red, green, gray, and blue balls represent the Bi, Fe, O, La, Ni, and Ti atoms, respectively.\label{fig:struct} }
\end{figure}

We carried out first principle calculations to investigate the ferroelectric controlled metal-insulator transition in the superlattices. We set 6 layers of BiFeO$_3$ in one supercell of (BiFeO$_3$)$_N$/LaNiO$_3$/LaTiO$_3$ ($N$=6) in our models as shown in Fig. ~\ref{fig:struct}. The spontaneous polarization of the bulk BiFeO$_3$ is along the [111] direction of the pseudo-cubic unit. With the compressive strain, the in-plane polarization tends to be suppressed so that the total polarization rotates towards [001]. When the in-plane lattice constant is about 3.71 \AA, the BiFeO$_3$ only films tend to transform to a ``$T$-like" phase with a large $c/a$ ratio about 1.2 to 1.3 and a large polarization approximately along [001]~\cite{PhysRevLett.102.217603,PhysRevB.81.054109}. However, in the  (BiFeO$_3$)$_6$/LaNiO$_3$/LaTiO$_3$ structure, the cooperative shift of the Bi ions along the out-of-plane direction is impeded by the LaNiO$_3$ and LaTiO$_3$ layers. Thus, the BiFeO$_3$ layers are in the ``R-like" phase with the $c/a$ ratio of about 1.1 and the out-of-plane polarization of about 60 $\mu$C/cm$^2$, which is alike to the BiFeO$_3$ films with in-plane lattice constants larger than 3.71 \AA. Here we only make use of the out-of-plane component of the polarization, because the only change of the structure by reversing of the in-plane component is a rotation around the out-of-plane axis and there is no difference in the electronic structure. Therefore we refer the polarization to the out-of-plane component of it in this paper. We define the state which the ferroelectric polarization orientation pointing from LaNiO$_3$ to LaTiO$_3$ as +P, while the state with opposite polarization as -P. A $\sqrt{2}\times\sqrt{2}$ pseudo-cubic in-plane lattice was used so that the oxygen octahedra can rotate freely. We kept the $ab$ in-plane lattice parameters fixed as 3.71 \AA , which is the LDA value of the substrate of LaAlO$_3$, and relax all the other degrees of freedom of the structure. The G-type antiferromagnetic structure was found to be stable in the (BiFeO$_3$)$_6$/LaNiO$_3$/LaTiO$_3$ structure.

We performed the density functional theory (DFT) calculation based on the projected augmented wave~\cite{kresse1999ultrasoft} (PAW) and local spin density approximation (LSDA)~\cite{perdew1981self} +Hubbard $U$ method~\cite{dudarev1998electron} as implemented in the Vienna \emph{ab initio} simulation package (VASP)~\cite{kresse1996efficient}. The filled valence states include the orbitals $5d6s6p$, $3d4s$, $5p5d6s$, $3s3p34d4s$, $3s3p3d4s$, and $2s2p$ for Bi, Fe, La, Ni, Ti, and O, respectively. A plane-wave basis set with energy cutoff of 500 eV and a $5\times5\times2$ $\Gamma$-centered k-point grid were used. The structures were relaxed until the forces are less than 0.03 eV/\AA. Though the LSDA +$U$ method cannot well describe the paramagnetic and insulating ground state in bulk LaTiO$_3$ simultaneously, it can describe the LaNiO$_3$/LaTiO$_3$ structure~\cite{chen2013modifying}. Effective $U$(Ni)=0 eV, $U$(Ti)=4 eV, and $U$(Fe)=6 eV are used throughout the work we presented here unless otherwise stated. The $U$(Fe) is used only to make the band gap of BiFeO$_3$ close to the experiment value. By following Ref.~\cite{chen2013modifying}, $U$(Ti)=4 eV was used to better align the Ti and Ni bands. We also found that $U$(Ti) has no significant effect on the overall results, because the Ti $3d$ bands are higher than the Ni $3d$ bands and are thus almost empty no matter how large the $U$(Ti) is. Disa \emph{et al.}~\cite{disa2015orbital} used the $U$(Ni) of 0 in a similar structure of LaAlO$_3$/LaNiO$_3$/LaTiO$_3$ and they found good agreement with the experimental results. We also compared the results with a series of $U$(Ni) ranging from 0 to 6 eV to see the effect of the Coulomb repulsion.

Two sets of Maximally localized Wannier functions (MLWFs)~\cite{mostofi2008wannier90} were constructed to investigate the mechanism of the metal-insulator transition using the Wannier90 package and its interface to VASP~\cite{franchini2012maximally}. The first set of MLWFs is $d$ band only, each function is of hybridized transitional metal $3d$ orbital and the neighboring O $2p$ feature. The other set of WLWFs, which is constructed to calculate the layer resolved polarization, consists of all the occupied-orbital-like wave functions.

\section{Results and Discussion}
\begin{figure}[htb]
  \centering
  \includegraphics[width=0.48\textwidth]{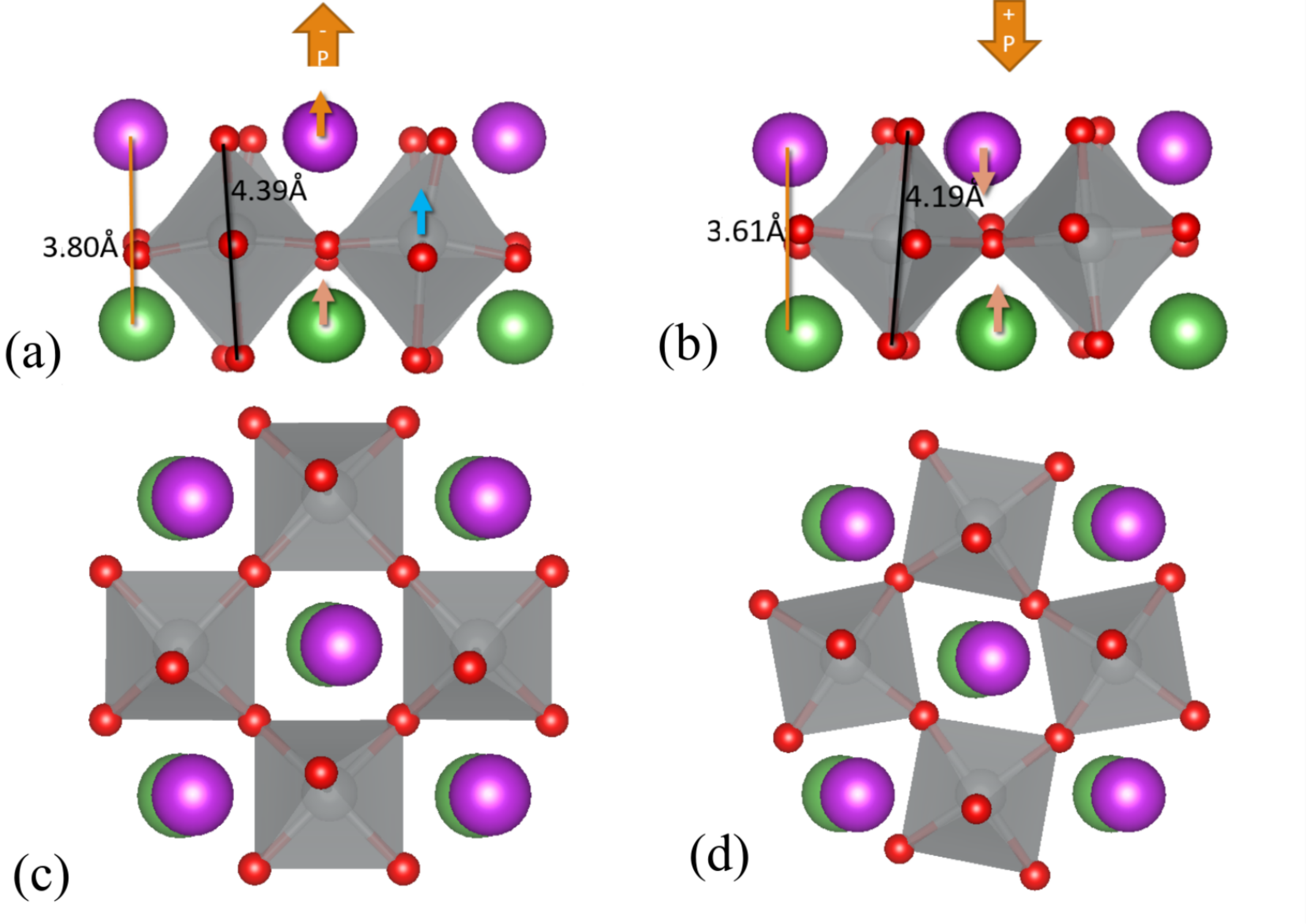}
  \caption{The structure near the LaNiO$_3$ layer. (a) The -P state, the a-c plane view. The brown arrows on the Bi and La atoms show the shifts of them. The blue arrow shows the shift of the Ni atom. (b) The +P state, the a-c plane view. (c) The -P state, the a-b plane view. (d) The +P state, the a-b plane view. The red arrows show the rotation of the oxygen octahedron. \label{fig:structsLNO} }
\end{figure}

In the (BiFeO$_3$)$_N$/LaNiO$_3$/LaTiO$_3$ structure, the bands near the Fermi energy are $e_g$ bands of Ni. We first look into the local structural distortions near the LaNiO$_3$ layer as shown in Fig.~\ref{fig:structsLNO}. The charge transfer between the LaTiO$_3$ and LaNiO$_3$ stays the same as in LaNiO$_3$/LaTiO$_3$ superlattices. La ions between NiO and TiO planes move towards the NiO plane to compensate the dipoles caused by the charge transfer. Bi ions at the BiFeO$_3$/LaNiO$_3$ interface shift away from the NiO plane in the -P state because of the ferroelectric polarization, while they shift towards the Ni-O plane in the +P state. 
As compared to the +P state, the distances between the Bi and La ions are larger in the -P state, and the diagonals of the oxygen octahedra along the out-of-plane direction in the frame of Bi and La ions are also longer; the in-plane Ni-O bond lengths are reduced from about 1.90 \AA~ in the +P state to about 1.88 \AA~ in the -P state to compensate the increasing of the volume of the oxygen octahedra caused by the elongation in the out-of-plane direction. Relative to the +P state, the larger Jahn-Teller distortion of the octahdron shifts the energy of the Ni $d_{x^2-y^2}$ orbital up and that of the Ni $d_{3z^2-r^2}$ orbital down in the -P state. Thus, the band gaps of the +P and -P states are different.

\begin{figure}[htb]
  \centering
  \includegraphics[width=0.48\textwidth]{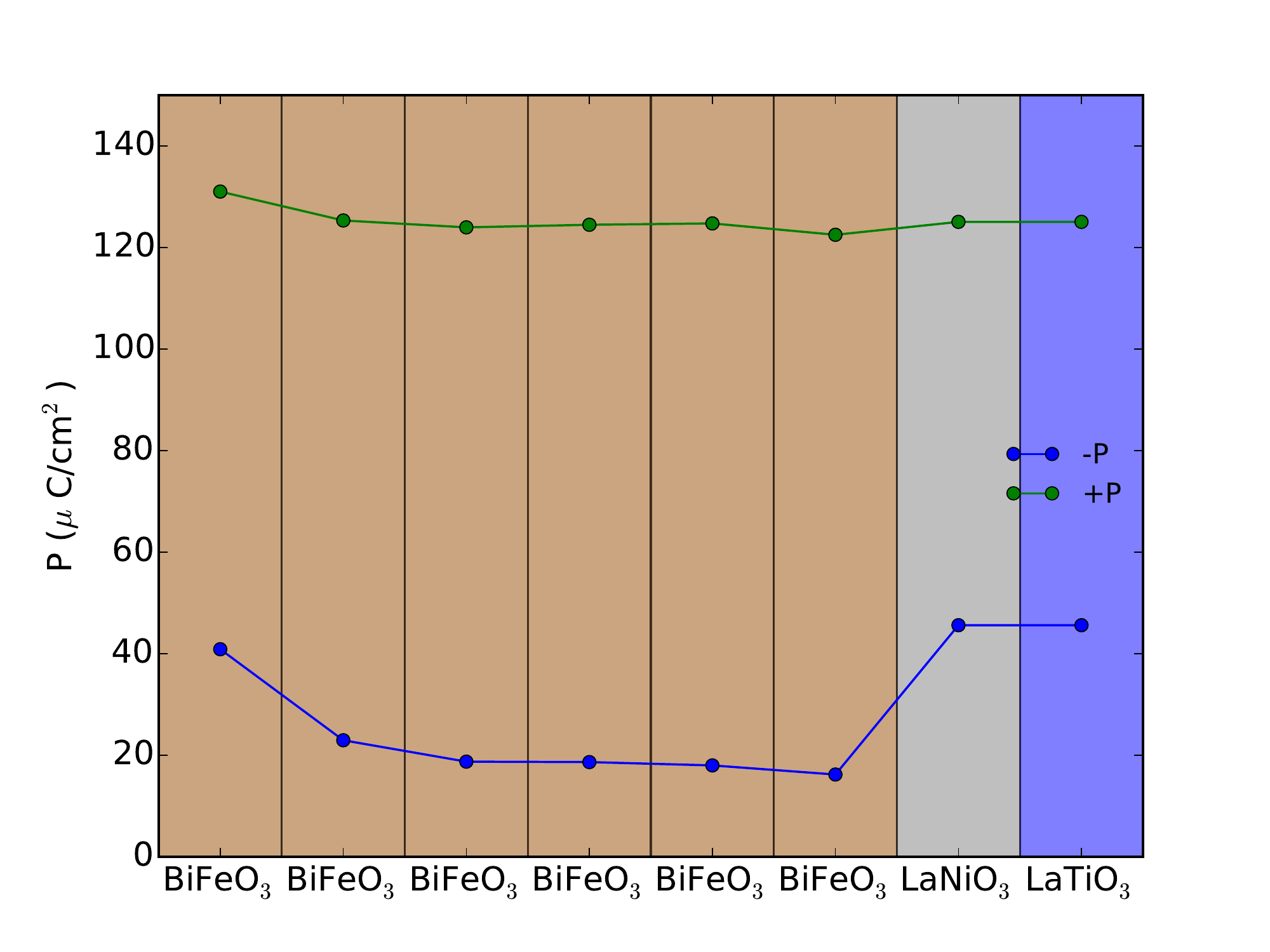}
  \caption{The layer resolved \emph{formal} polarization. The polarization in the LaNiO$_3$ and LaTiO$_3$ layers are the same because an averaged value of them are used. The charge neutral condition is not satisfied in these two layers due to the charge transfer between them. \label{fig:pcont} }
\end{figure}

The shorter in-plane Ni-O bonds in the -P state requires the Ni-O-Ni bond angles to be increased to accommodate the lattice constant fixed to the substrate.The details of the distortion and rotation patterns of the Ni-O octahedra are shown in Fig.~\ref{fig:structsLNO}. In the -P state, the rotations of the octahdedra are suppressed with the rotation angles less than 1$^\circ$. The directions of the ferroelectric polarization and the polarization caused by the LaNiO$_3$/LaTiO$_3$ charge transfer are opposite, leading to a polar discontinuity as shown in Fig.~\ref{fig:pcont}. The Ni ions shift relative to the O ions to compensate the polar discontinuity, thus the the O-Ni-O bond angles along the x (or y) direction are about 164$^\circ$ in the Ni-O octahedra. The oxygen octahedra in the tri-color superlattice rotate and tilt with the pattern $a^-b^-c^-$ in Glazer notation\cite{Glazer:a09401}. The oxygen octahedra tilt clockwise and anticlockwise alternately, with the tilting angles about 8$^\circ$. So the Ni-O octahedra in a chain are connected in clockwise-anticlockwise and anticlockwise-clockwise mode alternately. This shifting of Ni ions increase the Ni-O-Ni bond angles at the clockwise-anticlockwise connection, while decrease the angle at the anticlockwise-clockwise connection. Therefore, there are two kinds of Ni-O-Ni bond angles about 173$^\circ$ and 157$^\circ$, respectively. In the +P state, the rotation (with the angle about 10$^\circ$)  and tilting (with the angle about 6$^\circ$) of the octahedra are the predominant distortion patterns, which make the Ni-O-Ni bond angles about 157$^\circ$.

The Ni-O-Ni bond angles are larger in the -P state, thus the overlapping of the Ni $d_{x^2-y^2}$ and the O $2p$ orbitals is larger. What we can expect from this is a wider $d_{x^2-y^2}$ bandwidth and hopefully a metal-insulator transition.
\begin{figure}[t]
  \centering
  \includegraphics[width=0.38\textwidth]{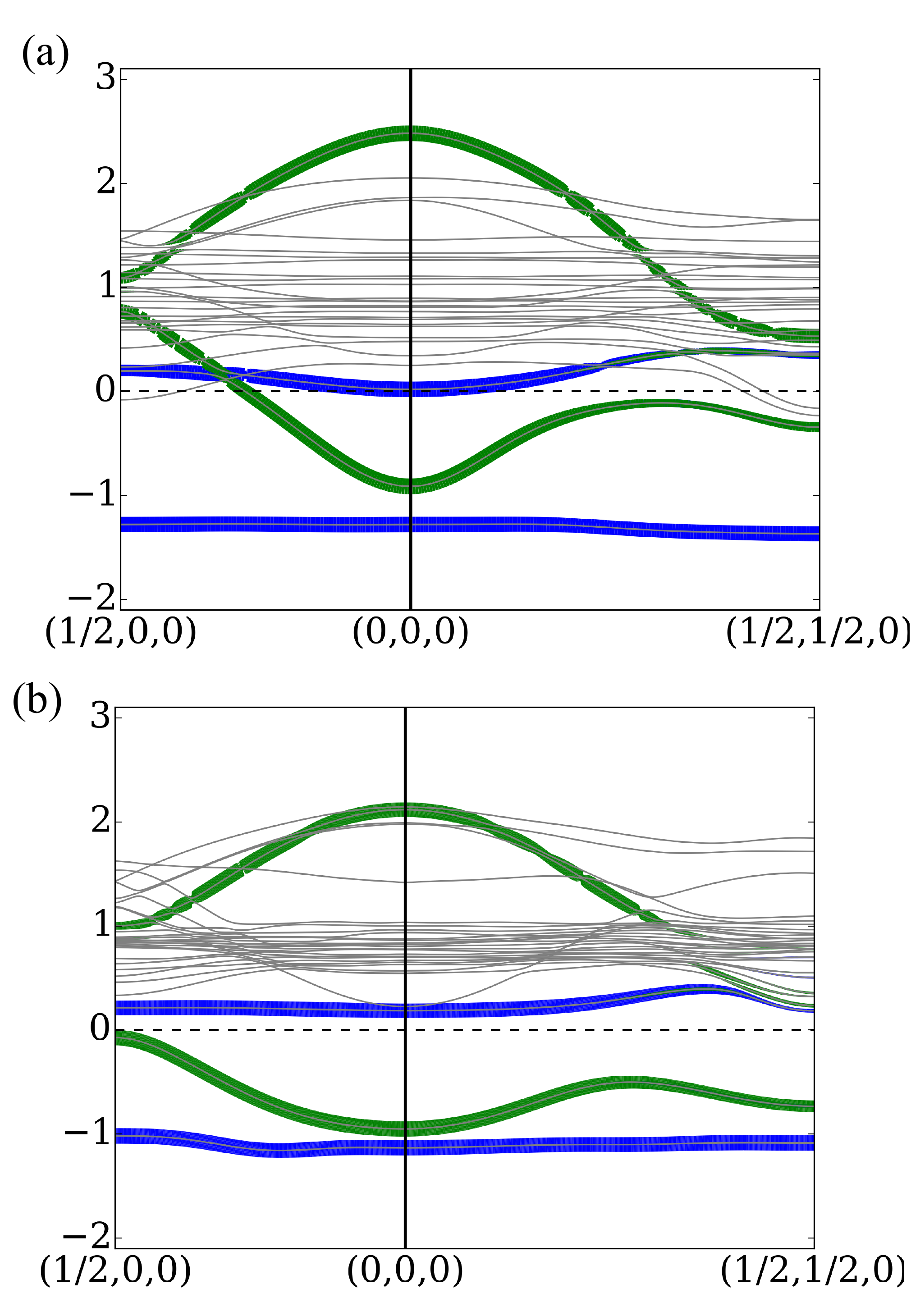}
  \caption{The band structures near the Fermi energy along the high symmetry lines of the Brillouin zone of the superlattice structure. (a) The -P state. (b) The +P state. The widths of the green and blue lines represent the projection onto the Ni $d_{x^2-y^2}$ and $d_{3z^2-r^2}$, respectively. \label{fig:Niband} }
\end{figure}

We calculated the band structure for the both the -P and +P states. The +P state has a band gap of about 0.2 eV, whereas the -P state is metallic. To see the detail of the change of the band structure, we constructed MLWFs and extracted an effective $d$ band only tight-binding Hamiltonian with the MLWFs as the basis set. Thus the MLWFs are of hybrid $d-p$ features. The band structure were calculated with this Hamiltonian, with one of the two spin channels (spin up) shown in Fig.\ref{fig:Niband}. The band structure of the other spin channel is the same because the structure is antiferromagnetic. The two Ni atoms in the calculated supercell with total spin up and down are denoted as Ni1 and Ni2 here, respectively. The two spin up $e_g$ states of Ni1 have lower energies than those of Ni2. The projection of the eigenstates to the Ni $e_g$ bands are shown. The four $e_g$ bands with spin up of the two Ni atoms are near the Fermi energies. It can be clearly seen that the metal-insulator transition is related to the change of the Ni $d_{x^2-y^2}$ bandwidths. The $d_{3z^2-r^2}$ bands are flat because the bands in the out-of-plane direction are confined. Whereas the in-plane $d_{x^2-y^2}$ bands can be seen as the bonding and antibonding states of the two Ni in-plane $e_g$ orbitals. The width of the band is proportional to the hopping integral $t$. 
 In the -P state, the hopping integrals are -0.41 eV and -0.36 eV corresponding to the Ni-O-Ni bonds with the angles 173$^\circ$ and 157$^\circ$, respectively. While in the +P state, the hopping integral is -0.37 eV corresponding to the Ni-O-Ni bonds with the angles 157$^\circ$. It can be seen that the $t\propto\cos^2\theta$ rule is well fitted. Therefore, the change of the bandwidths can be attributed to the structural distortion.

\begin{table}
\caption{The on-site energy of the Ni $e_g$ states. The energies are shifted so the Fermi energy is 0 eV. The unit is eV.\label{tab:onsite}}
\begin{center}
    \begin{tabular}{c c c c c}
      \hline
      \noalign{\vskip 1mm}
      &$d_{3z^2-r^2}^{Ni1}$ & $d_{x^2-y^2}^{Ni1}$ & $d_{3z^2-r^2}^{Ni2}$ & $d_{x^2-y^2}^{Ni2}$\\
      \hline
      \hline
      -P& -1.22& 0.53& 0.16&0.94 \\
      +P& -1.08&-0.15&0.16&0.8\\
\hline
    \end{tabular}
\end{center}
\end{table}
The relative shift of the Ni $e_g$ bands also largely affect the band gap. The on-site energies of the $e_g$ Wannier functions are listed in table~\ref{tab:onsite}. In the -P state, the crystal field splitting $\varepsilon(d_{x^2-y2})-\varepsilon(d_{3z^2-r^2})$ is larger than that that in the -P state, as the oxygen ocataheras are more elongated in the $z$ direction. Consequently, the on-site energy of $d_{x^2-y2}$ of Ni1 is raised even higher than that of the $d_{3z^2-r^2}$ of Ni2, leading to the crossover of the corresponding $d_{x^2-y2}$ and $d_{3z^2-r^2}$ bands.

We examined the possible effect of electron repulsion by adding an effective Hubbard $U(\text{Ni})$  ranging from 0 to 6 eV. The Hubbard $U$ pushes the occupied $e_g$ bands down and the unoccupied bands up, which inhibits the crossing over of these bands and the metallic state. With $U(\text{Ni})$ of above 4 eV, the metal-insulator transition is absent. Whether the metallic state can survive still needs experimental tests and better theoretical descriptions of the electron correlation effect. Nevertheless, the control of the bandwidths and band gap should be robust.

It should be noted that the carrier localization effect sometimes plays a role in 2D thin film structures, which make the structure with zero gap behave like insulators, like in Refs.~\cite{sakai2013gradual, PhysRevLett.106.246403}. We cannot rule out the possibility within our DFT calculation. Even so, a change in the conducting mechanism should be still observable.

From Fig.~\ref{fig:Niband}, the Ni $d_{x^2-y^2}$ a few Fe $t_{2g}$ bands at the BiFeO$_3$/LaTiO$_3$ interface cross the Fermi energy, showing a transfer of electrons from the LaNiO$_3$ to the BiFeO$_3$ layers. This can be understood as the result as the polarization discontinuity. The charge transfer from the LaTiO$_3$ to the LaNiO$_3$ layer caused a polarization pointing from LaNiO$_3$ to LaTiO$_3$, the orientation of which is the same as the ferroelectric polarization of BiFeO$_3$ in the +P state, while is different in the structure of -P. Though partly compensated by structural relaxation, the discontinuity can still lead to the charge transfer.

 We calculated the layer resolved polarization following the method of Stengel \emph{et al.} by constructing maximally localized Wannier functions~\cite{PhysRevB.80.241103,PhysRevB.83.235112}. The six Ni $t_{2g}$ and two $e_g$ electrons with majority spin, the five Fe $3d$ electrons in the majority spin channel, the O $2p$ electrons, and all the electrons at lower energies were considered as bounded charge. The charge neutral condition is not satisfied in the LaNiO$_3$ and LaTiO$_3$ layers because of the charge transfer between them. Therefore, we considered them as a whole and calculated the averaged polarization. As shown in Fig.~\ref{fig:pcont}, the polarizations are almost unified in the +P state, whereas the polarization discontinuity exists at both the BiFeO$_3$ interfaces in the -P state. According to the Polarization continuity condition $\frac{dP}{dx}=\rho$, the charge transfer from the LaNiO$_3$ layer to the BiFeO$_3$ layers. It should be noted that the polarizations in Fig.~\ref{fig:pcont} are the \emph{formal} polarization\cite{PhysRevB.80.241103}, not the \emph{effective} polarization which is commonly used to describe the spontaneous ferroelectric polarization. In the A$^{3+}$B$^{3+}$O$^{2-}_3$ perovskite structures, due to the (AO)$^+$(BO$_2$)$^-$ alignment, there is a component of the formal polarization of about $\frac{1}{2}|eR/\Omega|$ (about 70 $\mu$C/cm$^2$ in the BiFeO$_3$ layers), where $e$ is the electron charge, $R$ is the lattice constant along the out-of-plane direction of the pseudocubic, and $\Omega$ is the volume of the volume of the pseudocubic. By subtracting it from the formal polarizations, the residuals (which are approximately the \emph{effective} polarizations) in the BiFeO$_3$ layers are about +60 $\mu$C/cm$^2$ and -50 $\mu$C/cm$^2$ in the +P and -P states respectively.
\begin{figure}[h]
  \centering
  \includegraphics[width=0.4\textwidth]{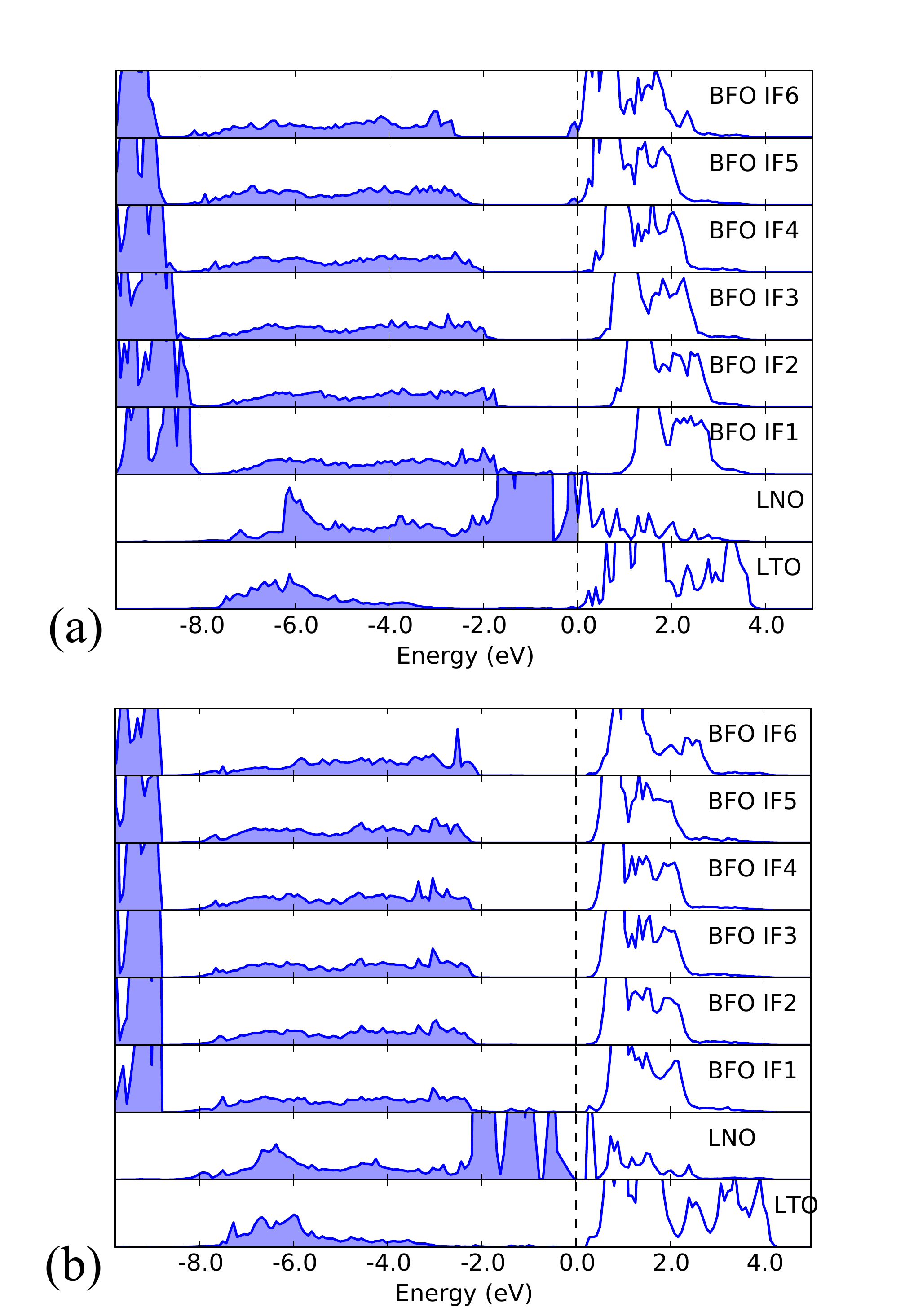}
  \caption{The band alignments of the (a) -P state and (b) +P state. The densities of states of the $3d$ electrons, which are near the Fermi energy, are shown. The BFO IFn means the n$^{\textrm{th}}$ layer from the BiFeO$_3$/LaNiO$_3$ interface.\label{fig:bandalign} }
\end{figure}
 We also calculated the band alignment of each layer, as shown in Fig.~\ref{fig:bandalign}. The polarization discontinuity causes the band shift in the BiFeO$_3$ in the -P state, whereas the much smaller polarization discontinuity causes almost no shift in the +P state. The result corroborated the charge transfer caused by the polarization discontinuity in -P state.

With the relative shift and the change of the width of the bands controlled by the reversing of the ferroelectric polarization, the Ni $e_g$ orbital polarization can also be modulated. The $e_g$ orbital polarization can be written as $P_{orb} = \frac{n_{3z^2-r^2}-n_{x^2-y^2}}{n_{3z^2-r^2}+n_{x^2-y^2}}$, where $n_{3z^2-r^2}$ and $n_{x^2-y^2}$ are the occupations of the $d_{3z^2-r^2}$ and $d_{x^2-y^2}$ orbitals, respectively, as opposite to that defined in Refs.~\cite{PhysRevB.82.134408,PhysRevLett.107.206804}. From Fig.~\ref{fig:Niband}, we can clearly see part of $d_{x^2-y^2}$ band which is beneath the Fermi energy in the +P state goes above the Fermi energy. $P_{orb}$ is 0.41 and 0.12 in the +P and -P state.

The magnetic properties of the LaNiO$_3$ can also be tuned. Because of the charge transfer and possibly the change of electron correlation effect which is related to the bandwidth, the local spin on Ni site is also tuned. In the -P state, the net spin projected on the Ni site is 0.66 $\mu_B$, while that in the +P state is 1.05 $\mu_B$.

\section{Conclusion}
In this work, we propose a strategy to control the electronic bandwidths and the band gap of a material stacked on the ferroelectric layer by reversing the ferroelectric polarization. This strategy is demonstrated to be feasible by designing a (BiFeO$_3$)$_N$/LaNiO$_3$/LaTiO$_3$ superlattice based on first principle calculations. The lattice distortion pattern changes with the orientation of the polarization. The on-site energies of the Ni $e_g$ orbitals related to the Jahn-Teller distortion of the Ni-O octahedron, and the widths of the in-plane Ni $d_{x^2-y^2}$ bands related the oxygen octahedron rotation angles and Ni-O-Ni bond angles, can both be thus be controlled by the reversing of the ferroelectric polarization. We also found the possibility of tuning of charge transfer between LaNiO$_3$ layers and the BiFeO$_3$ layers due to the change of polarization discontinuity condition. The $e_g$ orbital polarization and local Ni magnetic moment can also be manipulated.
\section*{Acknowledgement}
  The authors thank Jian-di Zhang for useful discussion. The work was supported by the National Basic Research Program of China (No. 2014CB921001), the National High Technology Research and Development Program of China (No. 2014AA032607), the National Natural Science Foundation of China (Nos. 11134012, 11404380, and 11474349), and the Strategic Priority Research Program (B) of the Chinese Academy of Sciences (No. XDB07030200).

\bibliographystyle{apsrev4-1}
\bibliography{mybib}

\end{document}